\documentclass{article}

 \usepackage[final]{neurips_2025_ml4ps}
 \usepackage{amsmath}
 \usepackage{graphicx}

\usepackage[utf8]{inputenc} 
\usepackage[T1]{fontenc}    
\usepackage{hyperref}       
\usepackage{url}            
\usepackage{booktabs}       
\usepackage{amsfonts}       
\usepackage{nicefrac}       
\usepackage{microtype}      
\usepackage{xcolor}         

\title{Reliable Parameter Inference for the Epoch of Reionization using Balanced Neural Ratio Estimation}

\author{%
  Diego González-Hernández \\
  Department of Physics\\
  University of California, Santa Barbara\\
  Santa Barbara, CA 93106, USA \\
  Lawrence Berkeley National Laboratory\\
  Berkeley, CA 94720, USA \\
  \texttt{dgonzalezhernandez@ucsb.edu} \\
  \And
  Molly Wolfson \\
  Center for Cosmology and AstroParticle Physics\\
  Department of Physics\\
  Department of Astronomy\\
  The Ohio State University\\
  Columbus, OH 43210, USA \\
  \texttt{wolfson.63@osu.edu} \\
  \And
  Joseph F. Hennawi \\
  Department of Physics\\
  University of California, Santa Barbara\\
  Santa Barbara, CA 93106, USA \\
  Leiden Observatory, Leiden University\\
  Niels Bohrweg 2, 2333 CA Leiden, Netherlands \\
  \texttt{joe@physics.ucsb.edu} \\
}

\begin{document}

\maketitle

\begin{abstract}
We present an application of the Balanced Neural Ratio Estimation (BNRE) algorithm to improve the statistical validity of parameter estimates used to characterize the Epoch of Reionization, where the common assumption of a multivariate Gaussian likelihood leads to overconfident and biased posterior distributions. Using a two-parameter model of the Ly$\alpha$ forest autocorrelation function, we show that BNRE yields posterior distributions that are significantly better calibrated than those obtained under the Gaussian likelihood assumption, as verified through the Test of Accuracy with Random Points (TARP) and Simulation-Based Calibration (SBC) diagnostics. These results demonstrate the potential of Simulation-Based Inference (SBI) methods, and in particular BNRE, to provide statistically robust parameter constraints within existing astrophysical modeling frameworks.
\end{abstract}

\section{Introduction}\label{sec:introduction}

The Epoch of Reionization (EoR) corresponds to the time in cosmic history when the neutral hydrogen in the intergalactic medium (IGM) was ionized by the first luminous sources \citep[see e.g.][]{Gnedin2022}. One of the primary probes of the late stages of the EoR is the Lyman-$\alpha$ (Ly$\alpha$) forest, a series of redshifted absorption features observed in the spectra of distant quasars caused by the presence of neutral hydrogen along the line of sight \citep{Gunn1965, Lynds1971}. Typically, two-point summary statistics of the Ly$\alpha$ forest (such as the one-dimensional power spectrum and the autocorrelation function) are used to understand the properties of the IGM during the EoR \citep[e.g.][]{Boera2019, Gaikwad2021, Walther2021, Wolfson2023thermal}. In standard analyses, parameter estimation relies on the assumption of a multivariate Gaussian likelihood for these statistics, an approximation that has been shown to yield overconfident or biased posteriors \citep[e.g.][]{Wolfson2023, Jin2025}. Such miscalibrations could lead to incorrect scientific conclusions, making it important to seek solutions that properly address this issue. Fortunately, the recent development of machine learning powered Simulation-Based Inference (SBI) methods has provided an alternative by directly leveraging forward simulations, allowing us to let go of explicit likelihood assumptions \citep{Cranmer2020}. In this work, we present our preliminary results obtained by applying Balanced Neural Ratio Estimation (BNRE) to a two-parameter model of the EoR, demonstrating its potential to improve the statistical validity in parameter inference problems within cosmology.\footnote{All code and analysis scripts are available at \url{https://github.com/diego-gonher/laf_sbi_bnre}.}

\section{Modeling the Epoch of Reionization}\label{sec:eor_models}
We adopt the model introduced in \citep{Wolfson2023}, which combines simulations of the IGM with semi-numerical calculations of the Ultraviolet Background (UVB). For our study, we focus on the redshift $z=5.5$, where this EoR model can be used to generate the mean autocorrelation function of the Ly$\alpha$ forest $\boldsymbol{\xi}_{m}$ as a function of two parameters: the mean free path of ionizing photons $\lambda_{\mathrm{mfp}}$, and the mean transmitted flux $\langle F \rangle$. We refer the reader to \citep{Wolfson2023} for specific details, but in summary, the model employs Nyx cosmological hydrodynamical simulations \citep{Almgren2013,Lukic2015} to model the underlying density, temperature, and velocity fields, while the method from \citep{Davies2016} is used to create spatially varying UVB realizations parameterized by $\lambda_{\mathrm{mfp}}$. Forward modeling is used to incorporate instrumental resolution, noise, and sightline lengths consistent with realistic observational data (e.g., the XQR-30 data from \citep{D'Odorico2023}). For each choice of $\boldsymbol{\theta} = \{\lambda_{\mathrm{mfp}}, \langle F \rangle\}$, the model produces ensembles of mock observations of the autocorrelation function of the Ly-$\alpha$ forest $\boldsymbol{\xi}_{i}$, the mean autocorrelation function $\boldsymbol{\xi}_{m}$ and the corresponding covariance matrix $\boldsymbol{\Sigma}_{\boldsymbol{\xi}}$.

A crucial aspect of this modeling approach is that each parameter combination is represented not by a single deterministic output, but by an arbitrarily large number of mock realizations required to properly sample the subsets of randomly selected sightlines and the observational noise. These mocks are already necessary for constructing model-dependent covariance matrices in traditional analyses, but they also naturally define a stochastic simulator. This makes the model ideally suited for Simulation-Based Inference (SBI), where the availability of many forward-modeled realizations per parameter point enables the training of neural estimators without additional modifications.

\section{Parameter Estimation}\label{sec:parameter_estimation}
Our goal is to infer the astrophysical parameters $\boldsymbol{\theta} = \{\lambda_{\mathrm{mfp}}, \langle F \rangle\}$ given an observed Ly$\alpha$ forest autocorrelation function $\boldsymbol{\xi}_{\mathrm{obs}}$. Using a Bayesian framework, the posterior distribution is given by:
\begin{equation}
p(\boldsymbol{\theta} \mid \boldsymbol{\xi}_{\mathrm{obs}}) \propto \mathcal{L}(\boldsymbol{\xi}_{\mathrm{obs}} \mid \boldsymbol{\theta}) \, p(\boldsymbol{\theta}),
\end{equation}
where $\mathcal{L}(\boldsymbol{\xi}_{\mathrm{obs}} \mid \boldsymbol{\theta})$ is the likelihood function and $p(\boldsymbol{\theta})$ is the prior over the model parameters. The choice of likelihood is therefore central to parameter inference, as it directly impacts the statistical validity of the resulting posterior distributions. 

We evaluate the statistical validity of the posterior distributions using two methods. First, we use \textit{coverage tests}, which assess whether the credible intervals of the inferred posterior distributions contain the true parameters with the expected frequency. Given our set of mock observations, coverage can be tested directly since we have access to their corresponding true $\boldsymbol{\theta}$. We employ the Test of Accuracy with Random Points \citep[TARP, ][]{Lemos2023}, which provides a diagnostic for posterior miscalibration by repeatedly comparing inferred credible regions against ground-truth parameter values. Second, we apply \textit{Simulation-Based Calibration} \citep[SBC, ][]{Cook2006, Talts2018}, which evaluates whether the ranks of true parameters within their corresponding posterior samples are uniformly distributed. While TARP quantifies global coverage performance, SBC provides a check of local miscalibration across the parameter space. Using both methods as inference tests provides a robust check on the statistical validity of the posteriors.

In what follows, we first describe the standard assumption of a multivariate Gaussian likelihood function adopted in previous works (Section~\ref{subsec:assumed-likelihood}), before describing an SBI method based on Balanced Neural Ratio Estimation (Section~\ref{subsec:bnre}).

\begin{figure*}
    \centering
    \includegraphics[width=1.\columnwidth]{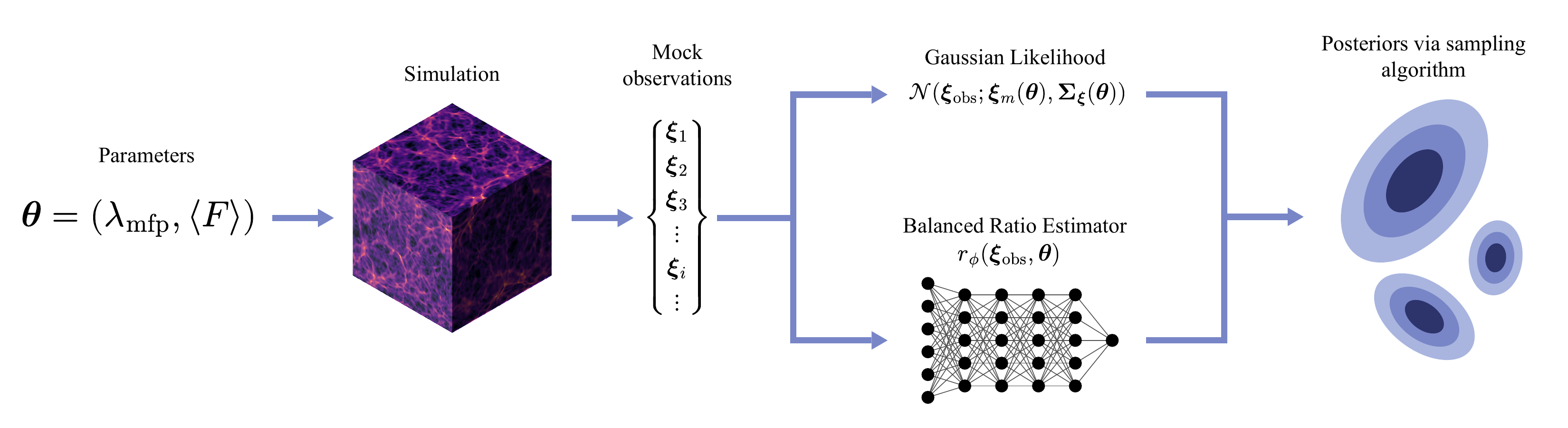}
    \caption{Given a set of parameters $\boldsymbol{\theta}$, the model generates a set of mock Ly$\alpha$ forest autocorrelation functions. These mocks can be used to: (i) compute the mean autocorrelation function $\boldsymbol{\xi}_{m}$ and the corresponding covariance matrix $\boldsymbol{\Sigma}_{\boldsymbol{\xi}}$ required to evaluate an assumed multivariate Gaussian likelihood, or (ii) train a neural ratio estimator (BNRE in this study). Both approaches yield posterior distributions via MCMC sampling (\texttt{emcee} in the Gaussian case, and HMC/NUTS in the BNRE case).}
    \label{fig:diagram}
\end{figure*}

\subsection{Assuming a Gaussian Likelihood}\label{subsec:assumed-likelihood}
A common approach in Ly$\alpha$ forest analyzes is to assume that the summary statistic of choice ($\boldsymbol{\xi}$ in this case) follows a multivariate Gaussian distribution with fixed model parameters. Under this assumption, the likelihood function can be written as:
\begin{equation}
\mathcal{L}(\boldsymbol{\xi}_{\mathrm{obs}} \mid \boldsymbol{\theta}) \propto
\exp\left[-\tfrac{1}{2}\left(\boldsymbol{\xi}_{\mathrm{obs}} - \boldsymbol{\xi}_{m}(\boldsymbol{\theta})\right)^{\top}
\boldsymbol{\Sigma}_{\boldsymbol{\xi}}(\boldsymbol{\theta})^{-1}
\left(\boldsymbol{\xi}_{\mathrm{obs}} - \boldsymbol{\xi}_{m}(\boldsymbol{\theta})\right)\right],
\end{equation}
Where $\boldsymbol{\xi}_{m}(\boldsymbol{\theta})$ and $\boldsymbol{\Sigma}_{\boldsymbol{\xi}}(\boldsymbol{\theta})$ denote the mean and covariance of the autocorrelation function evaluated at $\boldsymbol{\theta}$. In practice, these quantities are typically precomputed on a grid in parameter space, and inference is carried out using Markov Chain Monte Carlo (MCMC). Following \citep{Wolfson2023}, we use a grid of 557 distinct parameter combinations spanning $\boldsymbol{\theta} = \{\lambda_{\mathrm{mfp}}, \langle F \rangle\}$, with each grid point having its own $\boldsymbol{\xi}_{m}$ and $\boldsymbol{\Sigma}_{\boldsymbol{\xi}}$. We then use the \texttt{emcee} package \citep{ForemanMackey2013} with a nearest-grid-point interpolation scheme to evaluate $\boldsymbol{\xi}_{m}$ and $\boldsymbol{\Sigma}_{\boldsymbol{\xi}}$ between grid points. As mentioned above, this likelihood assumption often leads to overconfident or biased posteriors \citep[see Fig. 9 and Appendix C in][]{Wolfson2023}.

\subsection{Balanced Neural Ratio Estimation}\label{subsec:bnre}
Neural Ratio Estimation algorithms are a subset of SBI methods that bypass the need for an explicit likelihood function by training neural networks to approximate the ratio between the joint distribution $p(\boldsymbol{\xi}, \boldsymbol{\theta})$ and the product of marginals $p(\boldsymbol{\xi}) p(\boldsymbol{\theta})$ \citep{Cranmer2020}. Balanced Neural Ratio Estimation (BNRE) \citep{Delaunoy2022} extends this framework by introducing a tunable hyperparameter $\gamma$ in the loss function, which balances likelihood-ratio estimation accuracy against posterior calibration. By adjusting $\gamma$, BNRE can be made to nearly satisfy coverage tests.

We use the BNRE implementation in the \texttt{sbi} Python package \citep{Tejero-Cantero2020, Boelts2025}. The training dataset consists of $557 \times 500$ mock realizations $\boldsymbol{\xi}_i$, where 557 corresponds to the same $\boldsymbol{\theta}$ combinations on the grid described above (see Section~\ref{subsec:assumed-likelihood}), and for each $\boldsymbol{\theta}$ we have 500 distinct mocks (a random subset of the mocks used to estimate each $\boldsymbol{\Sigma}_{\boldsymbol{\xi}}$, which means no extra computations were required for the creation of this dataset). We use a 70–30 split for the training and validation sets and train the ratio estimator. The architecture for our estimator is the default ResNet-based classifier provided in the \texttt{sbi} package, which constructs a residual network with two blocks of 50 hidden units and ReLU activations, operating on the concatenated $(\boldsymbol{\xi}_i, \boldsymbol{\theta})$ inputs. For the training, we set $\gamma = 100$ (although we tested the effect of using $\gamma=\{10, 1000\}$, see Section~\ref{sec:conclusions} and Appendix~\ref{app:bnre_other_gammas}). Once trained, the ratio estimator defines the posterior via:
\vspace{0.2cm}
\begin{equation}
p(\boldsymbol{\theta} \mid \boldsymbol{\xi}_{\mathrm{obs}}) \propto r_{\phi}(\boldsymbol{\xi}_{\mathrm{obs}}, \boldsymbol{\theta}) \, p(\boldsymbol{\theta}),
\end{equation}

Where $r_{\phi}$ denotes the learned likelihood ratio. Because $r_{\phi}$ is differentiable with respect to $\boldsymbol{\theta}$, we can efficiently sample from the posterior using Hamiltonian Monte Carlo \citep[HMC,][]{Neal2011, Duane1987}. Specifically, we use the No-U-Turn Sampler \citep[NUTS,][]{Hoffman2011} implementation from the \texttt{pyro} library \citep{Bingham2019}, which exploits gradient information to explore parameter space more effectively. Consequently, the combination of BNRE with HMC is able to achieve both statistically valid and computationally efficient parameter estimations.

To summarize, both inference approaches rely on the same underlying dataset of mock autocorrelation functions, but use them in different ways: either to compute mean statistics and covariances for a Gaussian likelihood, or to train a neural ratio estimator. A schematic overview of both inference methods is provided in Figure~\ref{fig:diagram}.

\section{Results}\label{sec:results}
\begin{figure*}
    \centering
    \includegraphics[width=0.45\columnwidth]{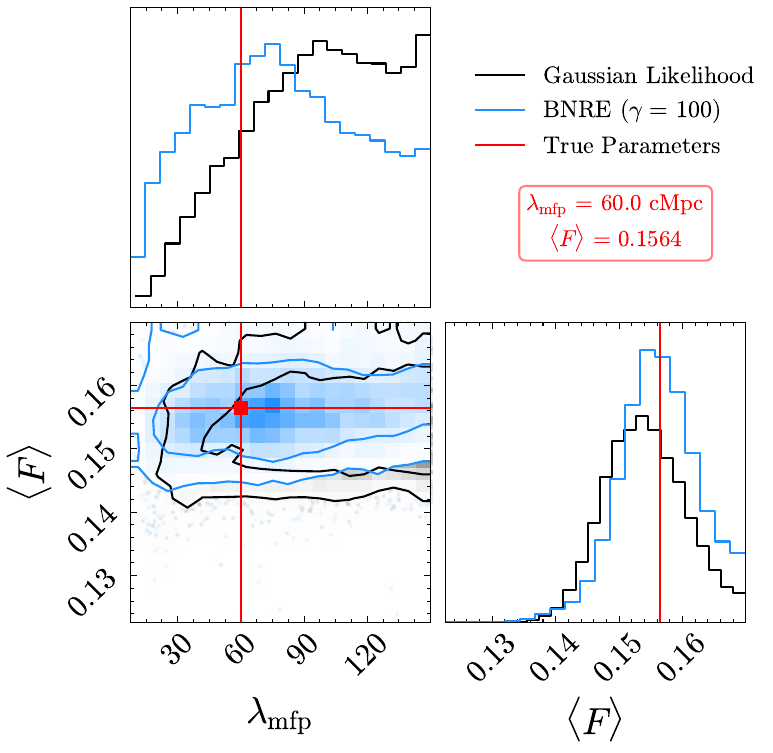}
    \hspace{0.07\textwidth} 
    \includegraphics[width=0.45\columnwidth]{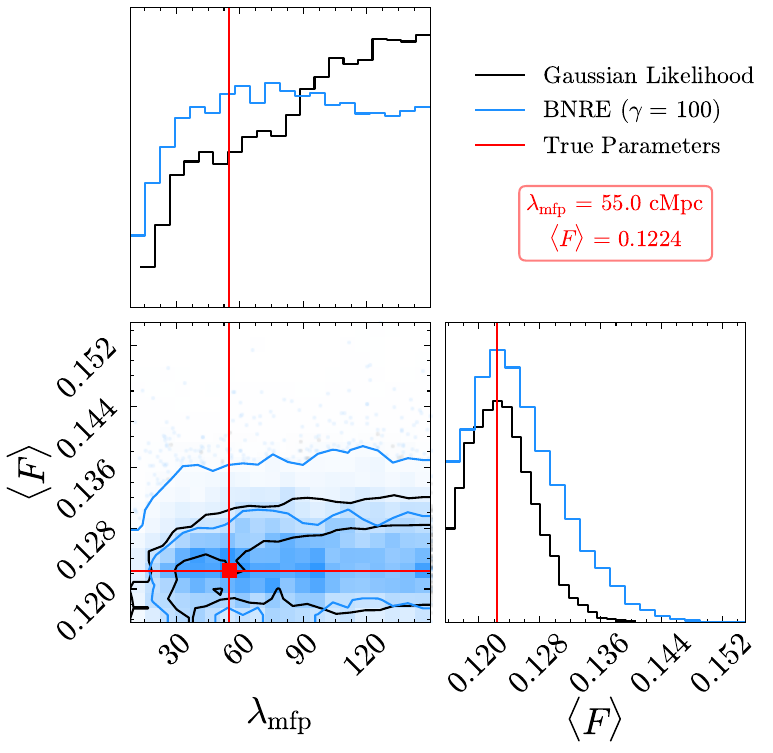}
    \caption{Corner plots of the posterior distributions obtained with both inference methods for two separate mock observations. The red text under the legend indicates the corresponding true parameter values $(\lambda_{\mathrm{mfp}}, \langle F \rangle)_{\mathrm{true}}$. Contours denote the 68\% and 95\% credible regions.}
    \label{fig:corner_plots}
\end{figure*}

To evaluate the performance of the two inference methods described above, we perform parameter inference on randomly selected mock observations (such that $\boldsymbol{\xi}_{\text{obs}}=\boldsymbol{\xi}_{i}$). For the MCMC sampling, we use four chains with 2000 samples each, discarding the first 1000 as ``burn-in''. This allows us to make a direct comparison of both inference methods, and test their statistical validity. As illustrative examples, Figure~\ref{fig:corner_plots} shows the posterior distributions obtained by either assuming a Gaussian likelihood or using our trained likelihood ratio estimator on two different mock observations, with additional examples shown in Figure~\ref{fig:extra_corner_plots} in Appendix~\ref{app:extra_corner_plots}. 

To quantitatively assess posterior calibration, we apply both parameter inference methods to a common set of 500 independent mock observations that were not used in training the BNRE's ratio estimator. The resulting posteriors are used to perform the inference tests presented below.

\subsection{Coverage Test}\label{subsec:coverage_test}
As mentioned in Section~\ref{sec:parameter_estimation}, we use TARP as our main coverage test. To estimate the statistical uncertainty in the coverage curves, we follow the bootstrapping procedure explained in \citep{Ruzza2025}. For both inference methods, we resample the set of mock observations and their associated posterior samples, recompute the TARP curve for each bootstrap realization, and use the resulting distribution to construct confidence intervals. 

The right panel of Figure~\ref{fig:inference_tests} shows the results of our coverage test. As can be seen, assuming a Gaussian likelihood leads to clearly overconfident or biased posterior distributions (shown in black). For the EoR model that we are using, this result is in agreement with the coverage test done in \citep{Wolfson2023}. In contrast, the coverage probability obtained with BNRE (shown in light blue) is close to the ideal curve, demonstrating that the posterior distributions obtained by this method are significantly less overconfident or biased. The corresponding shaded bands show the 16th--84th percentile ranges for both methods across 100 bootstrap replicates. 

\subsection{Simulation Based Calibration}\label{subsec:sbc}
\begin{figure*}
    \centering
    \includegraphics[width=0.48\columnwidth]{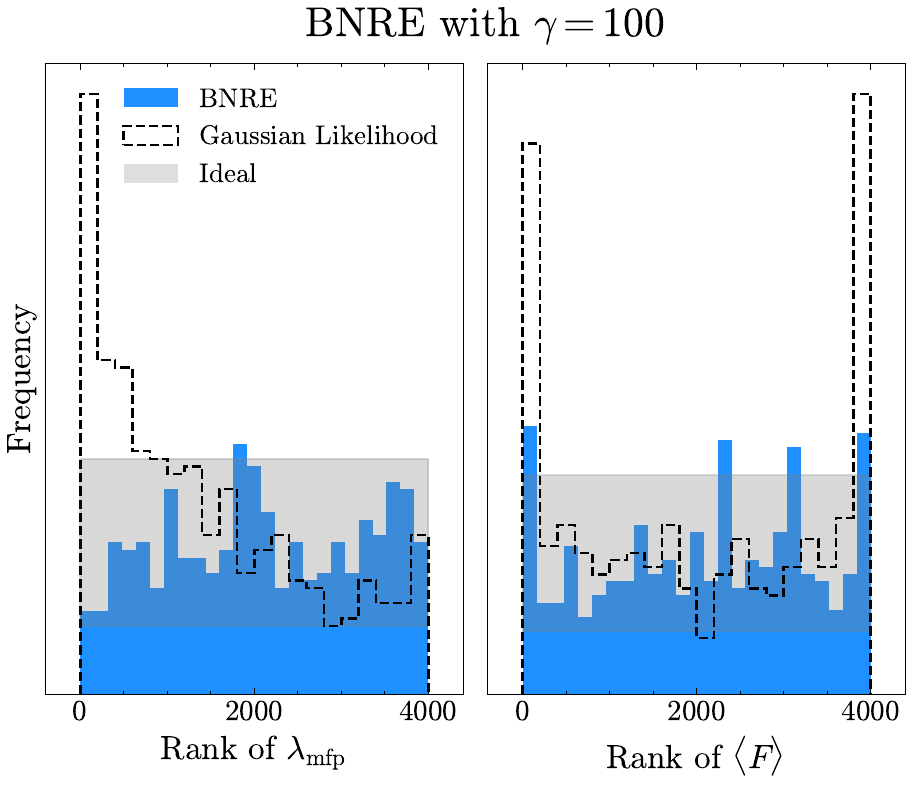}
    \hspace{0.07\textwidth} 
    \includegraphics[width=0.41\columnwidth]{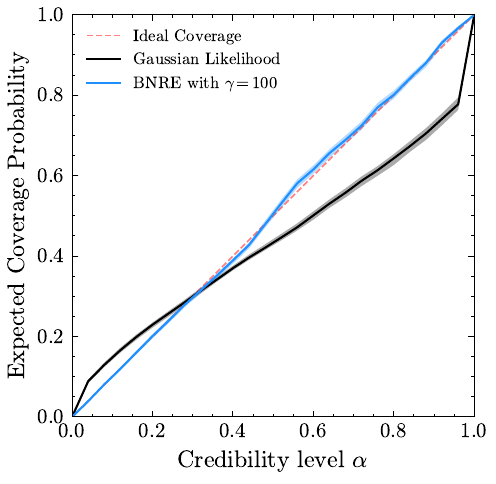}
    \caption{\textit{Left:} SBC rank histograms for $\lambda_{\text{mfp}}$ and $\langle F \rangle$ obtained by both parameter inference methods. Uniform posteriors indicate correct calibration, with the shaded region showing the expected range under sampling variability. \textit{Right:} Coverage probabilities obtained by using TARP on a set of posterior distributions obtained with both methods. The shaded regions represent the respective 16th--84th percentile ranges obtained via bootstrap sampling.}
    \label{fig:inference_tests}
\end{figure*}

As described in Section~\ref{sec:parameter_estimation}, we also perform SBC to further verify posterior validity. The left panel of Figure~\ref{fig:inference_tests} shows the SBC rank histograms computed using all posteriors obtained by both methods, with the shaded region indicating the expected variation under perfect uniformity \citep{Cook2006, Talts2018}. Assuming a Gaussian likelihood produces a skewed rank distribution for $\lambda_{\mathrm{mfp}}$ and a slightly U-shaped distribution for $\langle F \rangle$, indicating that this approach leads to biased posteriors for $\lambda_{\mathrm{mfp}}$ and slight overconfidence for $\langle F \rangle$. These results are consistent with the shape of the corresponding coverage curve. In contrast, the BNRE rank distributions for both parameters lie largely within the ideal uniformity region.

\section{Conclusions}\label{sec:conclusions}
Our results provide encouraging evidence that the BNRE algorithm can improve the statistical validity of parameter inference in cosmological models of the Epoch of Reionization. Although our study focuses on a two-parameter model, it is representative of common Ly$\alpha$ forest studies and serves as a non-trivial test for evaluating different inference methods. As mentioned in Section~\ref{subsec:bnre}, we also train the same BNRE architecture with $\gamma = \{10, 1000\}$ to examine the sensitivity of our results to this hyperparameter. Despite minor variations in the rank distributions of $\langle F \rangle$ (see Figures~\ref{fig:inference_tests_gamma_10} and~\ref{fig:inference_tests_gamma_1000} in Appendix~\ref{app:bnre_other_gammas}), all trained ratio estimators yield posterior distributions that are substantially better calibrated than those obtained under the Gaussian likelihood assumption.

To obtain a simple performance comparison between BNRE and a different SBI algorithm, we also train a Neural Posterior Estimation (NPE) and evaluate its performance (see Appendix~\ref{app:npe} for details). The resulting posteriors are clearly overconfident. We emphasize that this result is not representative of NPE’s overall potential. We expect that with adequate hyperparameter tuning and the use of a balanced objective \citep[as seen in][]{Delaunoy2023}, NPE could likely achieve a comparable performance. Nonetheless, training the NPE density estimator is substantially more computationally demanding than training the BNRE classifier, which constitutes a practical advantage for BNRE\footnote{In this study, training required approximately 7.9~CPU~hours for NPE and 1.2~CPU~hours for BNRE.}

In future work, we will explore the application of BNRE and other SBI methods to more complex, higher-dimensional reionization models. Importantly, standard modeling techniques across different subfields of astrophysics require little to no modifications to adopt SBI methods, as the same mock observations used to test inference pipelines can be repurposed for training different SBI algorithms. Our results therefore highlight how SBI methods such as BNRE can serve as a practical and reliable tool for enabling statistically robust scientific inference.

\newpage

\section*{Acknowledgments}
DGH acknowledges support for this work from NASA FINESST (Future Investigators in NASA Earth and Space Science and Technology) grant 80NSSC25K0313. JFH acknowledges support from the European Research Council (ERC) under the European Union’s Horizon 2020 research and innovation program (grant agreement No 885301) and from the National Science Foundation under Grant No. 1816006. This research also used resources of the National Energy Research Scientific Computing Center (NERSC), a U.S. Department of Energy Office of Science User Facility located at Lawrence Berkeley National Laboratory, operated under Contract No. DE-AC02-05CH11231.


\bibliographystyle{unsrt}
\bibliography{references}

\begin{thebibliography}{10}

\bibitem{Gnedin2022}
Nickolay~Y. {Gnedin} and Piero {Madau}.
\newblock {Modeling cosmic reionization}.
\newblock {\em Living Reviews in Computational Astrophysics}, 8(1):3, December 2022.

\bibitem{Gunn1965}
James~E. {Gunn} and Bruce~A. {Peterson}.
\newblock {On the Density of Neutral Hydrogen in Intergalactic Space.}
\newblock {\em \apj}, 142:1633--1636, November 1965.

\bibitem{Lynds1971}
Roger {Lynds}.
\newblock {The Absorption-Line Spectrum of 4c 05.34}.
\newblock {\em \apjl}, 164:L73, March 1971.

\bibitem{Boera2019}
Elisa {Boera}, George~D. {Becker}, James~S. {Bolton}, and Fahad {Nasir}.
\newblock {Revealing Reionization with the Thermal History of the Intergalactic Medium: New Constraints from the Ly{\ensuremath{\alpha}} Flux Power Spectrum}.
\newblock {\em \apj}, 872(1):101, February 2019.

\bibitem{Gaikwad2021}
Prakash {Gaikwad}, Raghunathan {Srianand}, Martin~G. {Haehnelt}, and Tirthankar~Roy {Choudhury}.
\newblock {A consistent and robust measurement of the thermal state of the IGM at 2 {\ensuremath{\leq}} z {\ensuremath{\leq}} 4 from a large sample of Ly {\ensuremath{\alpha}} forest spectra: evidence for late and rapid He II reionization}.
\newblock {\em \mnras}, 506(3):4389--4412, September 2021.

\bibitem{Walther2021}
Michael {Walther}, Eric {Armengaud}, Corentin {Ravoux}, Nathalie {Palanque-Delabrouille}, Christophe {Y{\`e}che}, and Zarija {Luki{\'c}}.
\newblock {Simulating intergalactic gas for DESI-like small scale Lyman{\ensuremath{\alpha}} forest observations}.
\newblock {\em \jcap}, 2021(4):059, April 2021.

\bibitem{Wolfson2023thermal}
Molly {Wolfson}, Joseph~F. {Hennawi}, Frederick~B. {Davies}, Zarija {Luki{\'c}}, and Jose {O{\~n}orbe}.
\newblock {Forecasting constraints on the high-z IGM thermal state from the Lyman-$\alpha$ forest flux auto-correlation function}.
\newblock {\em arXiv e-prints}, page arXiv:2309.05647, September 2023.

\bibitem{Wolfson2023}
Molly {Wolfson}, Joseph~F. {Hennawi}, Frederick~B. {Davies}, and Jose {O{\~n}orbe}.
\newblock {Forecasting constraints on the mean free path of ionizing photons at z {\ensuremath{\geq}} 5.4 from the Lyman-{\ensuremath{\alpha}} forest flux autocorrelation function}.
\newblock {\em \mnras}, 521(3):4056--4073, May 2023.

\bibitem{Jin2025}
Zhenyu {Jin}, Molly {Wolfson}, Joseph~F. {Hennawi}, and Diego {Gonz{\'a}lez-Hern{\'a}ndez}.
\newblock {Neural network emulator to constrain the high-z IGM thermal state from Lyman-{\ensuremath{\alpha}} forest flux autocorrelation function}.
\newblock {\em \mnras}, 536(3):2277--2293, January 2025.

\bibitem{Cranmer2020}
Kyle {Cranmer}, Johann {Brehmer}, and Gilles {Louppe}.
\newblock {The frontier of simulation-based inference}.
\newblock {\em Proceedings of the National Academy of Science}, 117(48):30055--30062, December 2020.

\bibitem{Almgren2013}
Ann~S. {Almgren}, John~B. {Bell}, Mike~J. {Lijewski}, Zarija {Luki{\'c}}, and Ethan {Van Andel}.
\newblock {Nyx: A Massively Parallel AMR Code for Computational Cosmology}.
\newblock {\em \apj}, 765(1):39, March 2013.

\bibitem{Lukic2015}
Zarija {Luki{\'c}}, Casey~W. {Stark}, Peter {Nugent}, Martin {White}, Avery~A. {Meiksin}, and Ann {Almgren}.
\newblock {The Lyman {\ensuremath{\alpha}} forest in optically thin hydrodynamical simulations}.
\newblock {\em \mnras}, 446(4):3697--3724, February 2015.

\bibitem{Davies2016}
Frederick~B. {Davies} and Steven~R. {Furlanetto}.
\newblock {Large fluctuations in the hydrogen-ionizing background and mean free path following the epoch of reionization}.
\newblock {\em \mnras}, 460(2):1328--1339, August 2016.

\bibitem{D'Odorico2023}
Valentina {D'Odorico}, E.~{Ba{\~n}ados}, G.~D. {Becker}, M.~{Bischetti}, S.~E.~I. {Bosman}, G.~{Cupani}, R.~{Davies}, E.~P. {Farina}, A.~{Ferrara}, C.~{Feruglio}, C.~{Mazzucchelli}, E.~{Ryan-Weber}, J.~T. {Schindler}, A.~{Sodini}, B.~P. {Venemans}, F.~{Walter}, H.~{Chen}, S.~{Lai}, Y.~{Zhu}, F.~{Bian}, S.~{Campo}, S.~{Carniani}, S.~{Cristiani}, F.~{Davies}, R.~{Decarli}, A.~{Drake}, A.~C. {Eilers}, X.~{Fan}, P.~{Gaikwad}, S.~{Gallerani}, B.~{Greig}, M.~G. {Haehnelt}, J.~{Hennawi}, L.~{Keating}, G.~{Kulkarni}, A.~{Mesinger}, R.~A. {Meyer}, M.~{Neeleman}, M.~{Onoue}, A.~{Pallottini}, Y.~{Qin}, S.~{Rojas-Ruiz}, S.~{Satyavolu}, A.~{Sebastian}, R.~{Tripodi}, F.~{Wang}, M.~{Wolfson}, J.~{Yang}, and M.~V. {Zanchettin}.
\newblock {XQR-30: The ultimate XSHOOTER quasar sample at the reionization epoch}.
\newblock {\em \mnras}, 523(1):1399--1420, July 2023.

\bibitem{Lemos2023}
Pablo {Lemos}, Adam {Coogan}, Yashar {Hezaveh}, and Laurence {Perreault-Levasseur}.
\newblock {Sampling-Based Accuracy Testing of Posterior Estimators for General Inference}.
\newblock {\em 40th International Conference on Machine Learning}, 202:19256--19273, January 2023.

\bibitem{Cook2006}
Samantha~R Cook, Andrew Gelman, and Donald~B Rubin.
\newblock Validation of software for bayesian models using posterior quantiles.
\newblock {\em Journal of Computational and Graphical Statistics}, 15(3):675--692, 2006.

\bibitem{Talts2018}
Sean {Talts}, Michael {Betancourt}, Daniel {Simpson}, Aki {Vehtari}, and Andrew {Gelman}.
\newblock {Validating Bayesian Inference Algorithms with Simulation-Based Calibration}.
\newblock {\em arXiv e-prints}, page arXiv:1804.06788, April 2018.

\bibitem{ForemanMackey2013}
Daniel {Foreman-Mackey}, David~W. {Hogg}, Dustin {Lang}, and Jonathan {Goodman}.
\newblock {emcee: The MCMC Hammer}.
\newblock {\em \pasp}, 125(925):306, March 2013.

\bibitem{Delaunoy2022}
Arnaud {Delaunoy}, Joeri {Hermans}, Fran{\c{c}}ois {Rozet}, Antoine {Wehenkel}, and Gilles {Louppe}.
\newblock {Towards Reliable Simulation-Based Inference with Balanced Neural Ratio Estimation}.
\newblock {\em arXiv e-prints}, page arXiv:2208.13624, August 2022.

\bibitem{Tejero-Cantero2020}
Alvaro Tejero-Cantero, Jan Boelts, Michael Deistler, Jan-Matthis Lueckmann, Conor Durkan, Pedro~J. Gonçalves, David~S. Greenberg, and Jakob~H. Macke.
\newblock sbi: A toolkit for simulation-based inference.
\newblock {\em Journal of Open Source Software}, 5(52):2505, 2020.

\bibitem{Boelts2025}
Jan {Boelts}, Michael {Deistler}, Manuel {Gloeckler}, {\'A}lvaro {Tejero-Cantero}, Jan-Matthis {Lueckmann}, Guy {Moss}, Peter {Steinbach}, Thomas {Moreau}, Fabio {Muratore}, Julia {Linhart}, Conor {Durkan}, Julius {Vetter}, Benjamin {Miller}, Maternus {Herold}, Abolfazl {Ziaeemehr}, Matthijs {Pals}, Theo {Gruner}, Sebastian {Bischoff}, Nastya {Krouglova}, Richard {Gao}, Janne {Lappalainen}, B{\'a}lint {Mucs{\'a}nyi}, Felix {Pei}, Auguste {Schulz}, Zinovia {Stefanidi}, Pedro {Rodrigues}, Cornelius {Schr{\"o}der}, Faried {Zaid}, Jonas {Beck}, Jaivardhan {Kapoor}, David {Greenberg}, Pedro {Gon{\c{c}}alves}, and Jakob {Macke}.
\newblock {sbi reloaded: a toolkit for simulation-based inference workflows}.
\newblock {\em The Journal of Open Source Software}, 10(108):7754, April 2025.

\bibitem{Neal2011}
Radford {Neal}.
\newblock {MCMC Using Hamiltonian Dynamics}.
\newblock In {\em Handbook of Markov Chain Monte Carlo}, pages 113--162. 2011.

\bibitem{Duane1987}
Simon {Duane}, A.D. {Kennedy}, Brian~J. {Pendleton}, and Duncan {Roweth}.
\newblock Hybrid monte carlo.
\newblock {\em Physics Letters B}, 195(2):216--222, 1987.

\bibitem{Hoffman2011}
Matthew~D. {Hoffman} and Andrew {Gelman}.
\newblock {The No-U-Turn Sampler: Adaptively Setting Path Lengths in Hamiltonian Monte Carlo}.
\newblock {\em arXiv e-prints}, page arXiv:1111.4246, November 2011.

\bibitem{Bingham2019}
Eli Bingham, Jonathan~P. Chen, Martin Jankowiak, Fritz Obermeyer, Neeraj Pradhan, Theofanis Karaletsos, Rohit Singh, Paul~A. Szerlip, Paul Horsfall, and Noah~D. Goodman.
\newblock Pyro: Deep universal probabilistic programming.
\newblock {\em J. Mach. Learn. Res.}, 20:28:1--28:6, 2019.

\bibitem{Ruzza2025}
A.~{Ruzza}, G.~{Lodato}, G.~P. {Rosotti}, and P.~J. {Armitage}.
\newblock {DBNets2.0: Simulation-based inference for planet-induced dust substructures in protoplanetary discs}.
\newblock {\em \aap}, 700:A190, August 2025.

\bibitem{Delaunoy2023}
Arnaud {Delaunoy}, Benjamin~Kurt {Miller}, Patrick {Forr{\'e}}, Christoph {Weniger}, and Gilles {Louppe}.
\newblock {Balancing Simulation-based Inference for Conservative Posteriors}.
\newblock {\em arXiv e-prints}, page arXiv:2304.10978, April 2023.

\bibitem{Greenberg2019}
David~S. {Greenberg}, Marcel {Nonnenmacher}, and Jakob~H. {Macke}.
\newblock {Automatic Posterior Transformation for Likelihood-Free Inference}.
\newblock {\em arXiv e-prints}, page arXiv:1905.07488, May 2019.

\bibitem{Hermans2021}
Joeri Hermans, Arnaud Delaunoy, Fran{\c{c}}ois Rozet, Antoine Wehenkel, Volodimir Begy, and Gilles Louppe.
\newblock A crisis in simulation-based inference? beware, your posterior approximations can be unfaithful.
\newblock {\em Transactions on Machine Learning Research}, 2022.

\end{thebibliography}


\appendix

\section{Additional Corner Plots}\label{app:extra_corner_plots}
\begin{figure*}[!htbp]
    \centering
    \begin{minipage}[t]{0.45\textwidth}
        \centering
        \includegraphics[width=\linewidth]{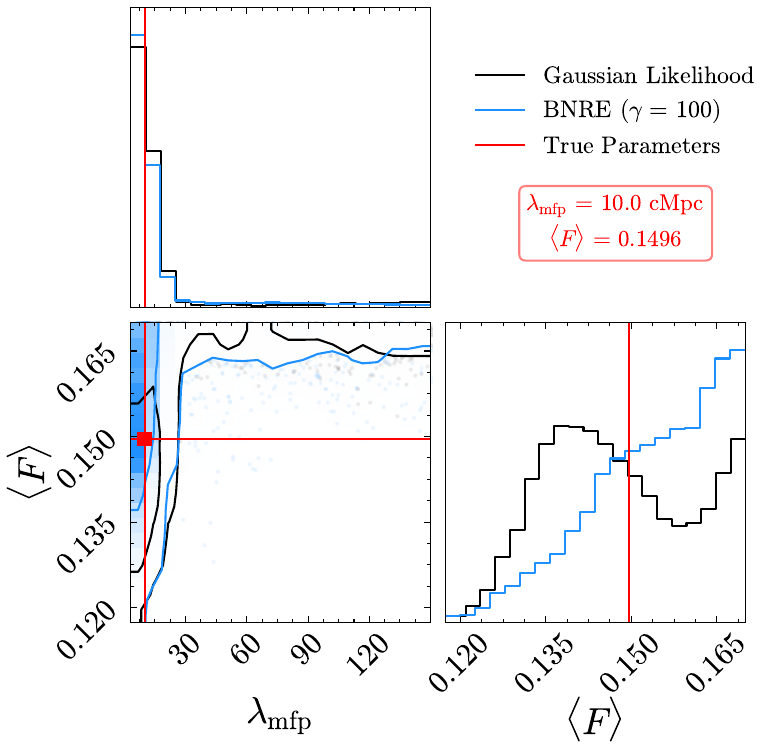}
    \end{minipage}
    \hspace{0.07\textwidth}
    \begin{minipage}[t]{0.45\textwidth}
        \centering
        \includegraphics[width=\linewidth]{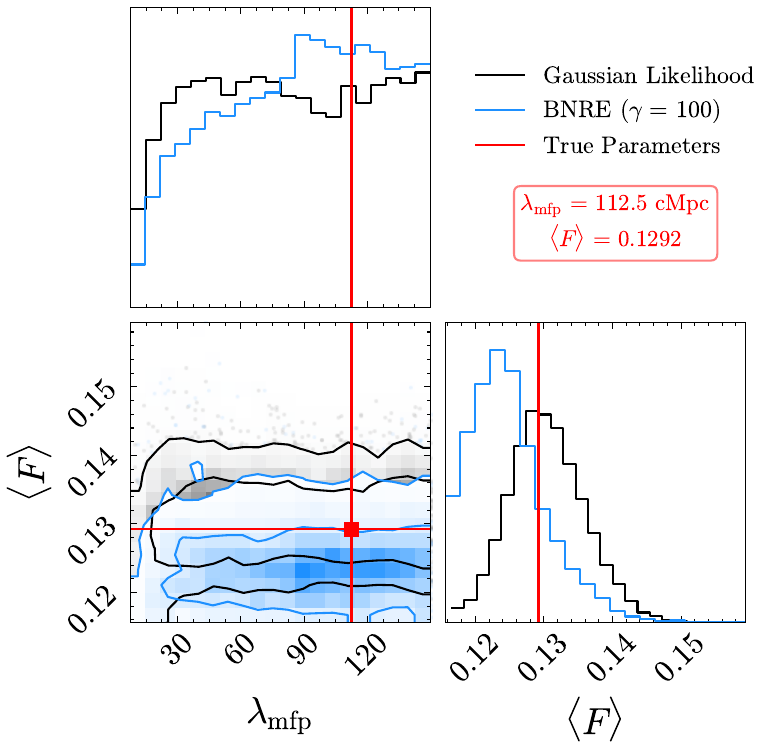}
    \end{minipage}

    \vspace{0.05\textwidth} 

    \begin{minipage}[t]{0.45\textwidth}
        \centering
        \includegraphics[width=\linewidth]{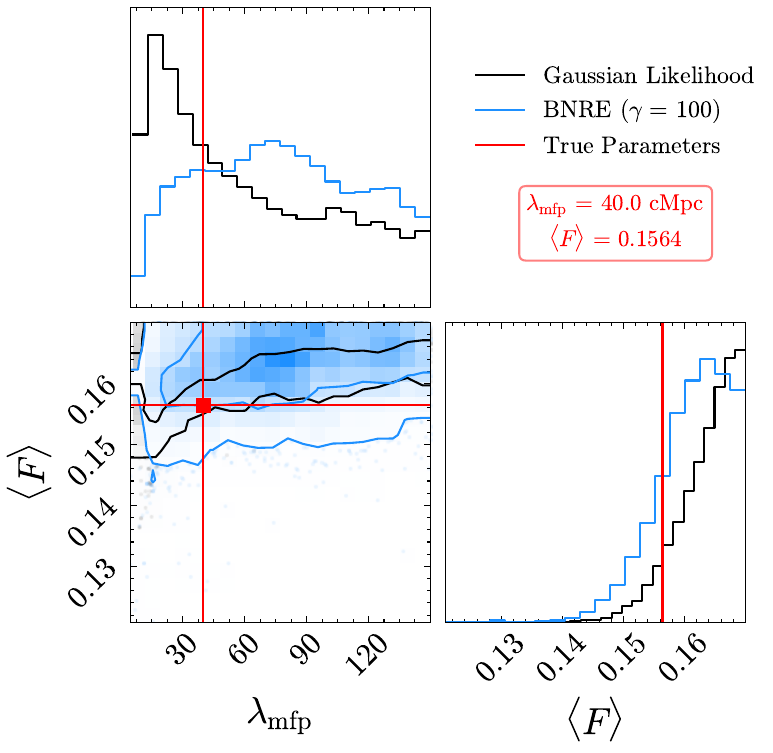}
    \end{minipage}
    \hspace{0.07\textwidth}
    \begin{minipage}[t]{0.45\textwidth}
        \centering
        \includegraphics[width=\linewidth]{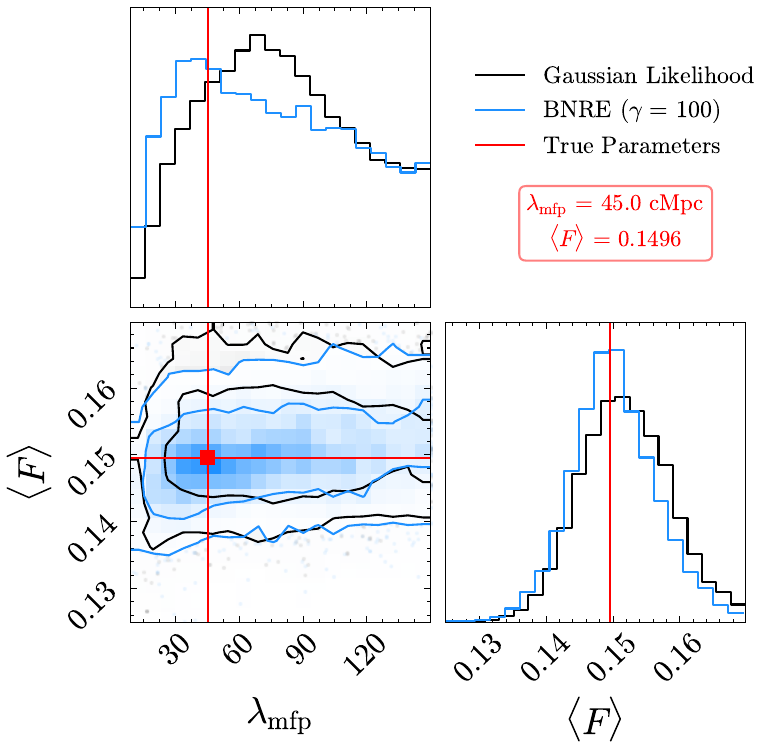}
    \end{minipage}

    \caption{Corner plots of the posterior distributions obtained using the Guassian likelihood and BNRE with $\gamma=100.0$ for four randomly selected mock observations. The true parameter values are shown in red for each case respectively. The contours denote the 68\% and 95\% credible regions.}
    \label{fig:extra_corner_plots}
\end{figure*}

Figure~\ref{fig:extra_corner_plots} presents four additional (randomly selected) examples comparing the posterior distributions obtained using the Gaussian likelihood and BNRE with $\gamma = 100.0$. These examples show the typical differences in posterior shape and coverage across independent mock observations.

\section{Performance of BNRE with $\gamma=10$ and $\gamma=1000$}\label{app:bnre_other_gammas}
\begin{figure*}
    \centering
    \includegraphics[width=0.48\columnwidth]{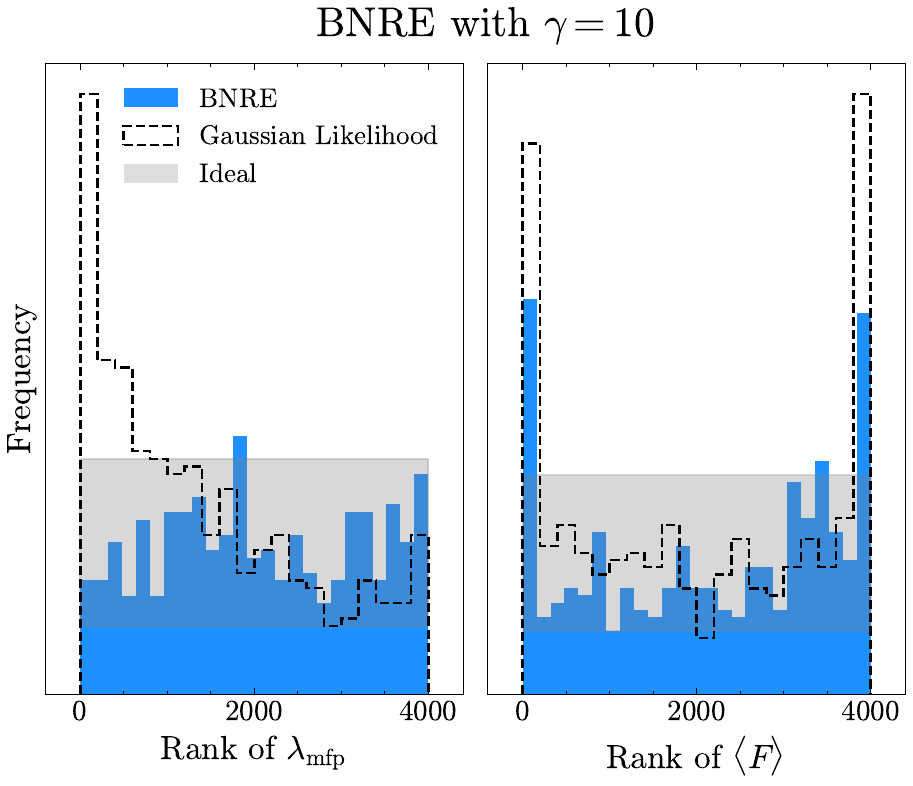}
    \hspace{0.07\textwidth} 
    \includegraphics[width=0.41\columnwidth]{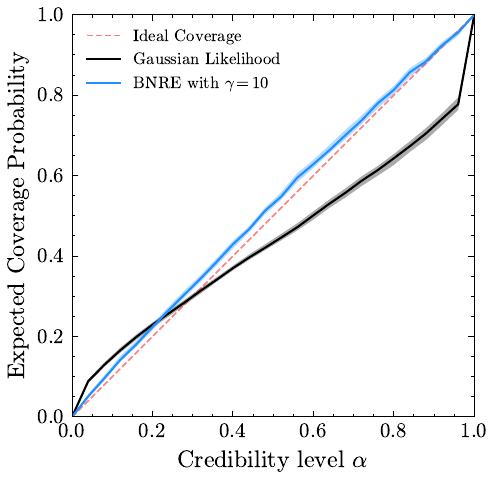}
    \caption{\textit{Left:} SBC rank histograms for $\lambda_{\text{mfp}}$ and $\langle F \rangle$ obtained using BNRE with $\gamma=10$. Uniform posteriors indicate correct calibration, with the shaded region showing the expected range under sampling variability. \textit{Right:} Coverage probabilities obtained by using TARP on a set of posterior distributions obtained with both methods. The shaded regions represent the respective 16th--84th percentile ranges obtained via bootstrap sampling.}
    \label{fig:inference_tests_gamma_10}
\end{figure*}

\begin{figure*}
    \centering
    \includegraphics[width=0.48\columnwidth]{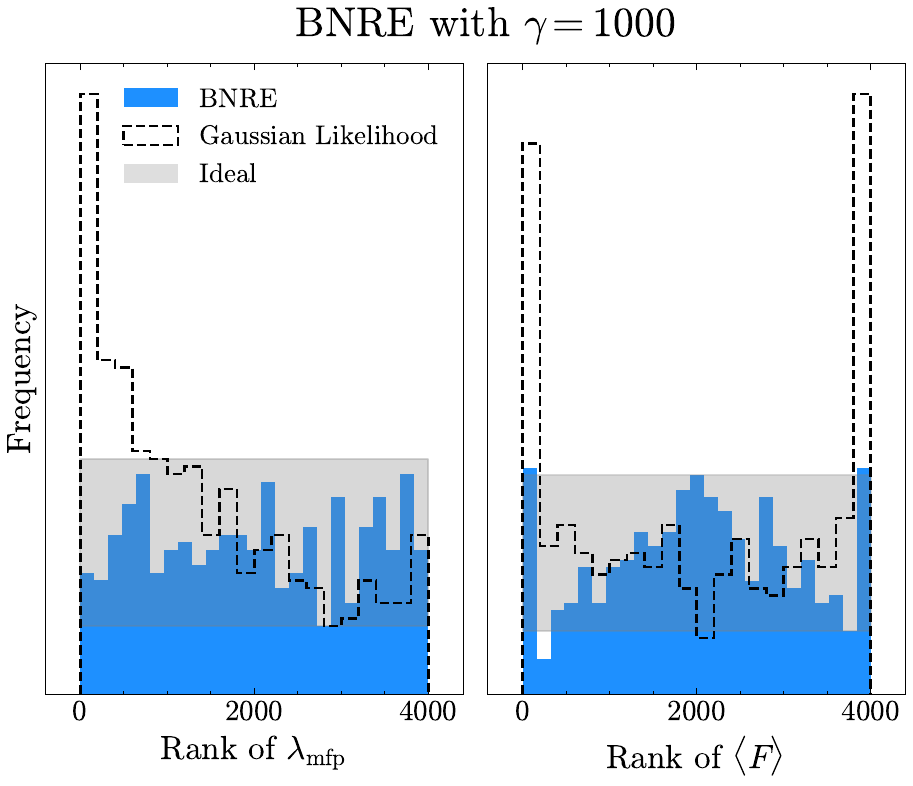}
    \hspace{0.07\textwidth} 
    \includegraphics[width=0.41\columnwidth]{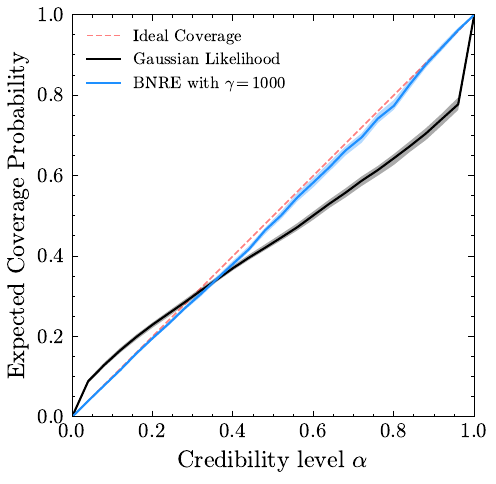}
    \caption{\textit{Left:} SBC rank histograms for $\lambda_{\text{mfp}}$ and $\langle F \rangle$ obtained using BNRE with $\gamma=1000$. Uniform posteriors indicate correct calibration, with the shaded region showing the expected range under sampling variability. \textit{Right:} Coverage probabilities obtained by using TARP on a set of posterior distributions obtained with both methods. The shaded regions represent the respective 16th--84th percentile ranges obtained via bootstrap sampling.}
    \label{fig:inference_tests_gamma_1000}
\end{figure*}

Using the same training set and architecture described in Section~\ref{subsec:bnre}, we train two additional ratio estimators with $\gamma = \{10, 1000\}$. Following the procedure outlined in Section~\ref{sec:results}, we perform parameter inference with both estimators to obtain two respective sets of 500 posterior distributions, and then apply TARP and SBC to directly compare their performance. The corresponding results are shown in Figures~\ref{fig:inference_tests_gamma_10} and~\ref{fig:inference_tests_gamma_1000}. 

Both models produce coverage curves that are close to ideal, demonstrating that the calibration of the posterior distributions is robust to moderate changes in $\gamma$. The SBC histograms show that the rank distributions for $\lambda_{\mathrm{mfp}}$ and $\langle F \rangle$ also remain largely within the ideal region. However, the rank distribution for $\langle F \rangle$ exhibits a slight U-shape for the $\gamma=10$ case and a slight inverted U-shape for $\gamma=1000$. This behavior is expected, as increasing $\gamma$ reduces the relative weight of the likelihood-ratio loss term, leading to less confident posteriors \citep[see][]{Delaunoy2022}.

\section{Neural Posterior Estimation}\label{app:npe}
To provide a simple comparison against an alternative simulation-based inference (SBI) method, we also train a Neural Posterior Estimation (NPE) algorithm. In NPE, a density network is used to model the posterior $p(\boldsymbol{\theta} \mid \boldsymbol{\xi})$ directly. We use the same training and validation datasets as for the BNRE networks (see Section~\ref{subsec:bnre}) to train a posterior estimator, employing the NPE method from \citep{Greenberg2019} available in the \texttt{sbi} package \citep{Tejero-Cantero2020}. 

We use the default configuration for the density estimator provided in \texttt{sbi}, which consists of a Masked Autoregressive Flow (MAF) composed of five affine autoregressive transforms with random permutations between them. Each transform uses fully connected layers with 50 hidden units and $\tanh$ activations. The model is optimized using maximum likelihood with early stopping based on the validation loss, again using the default settings in \texttt{sbi}. 

\begin{figure*}
    \centering
    \includegraphics[width=0.49\columnwidth]{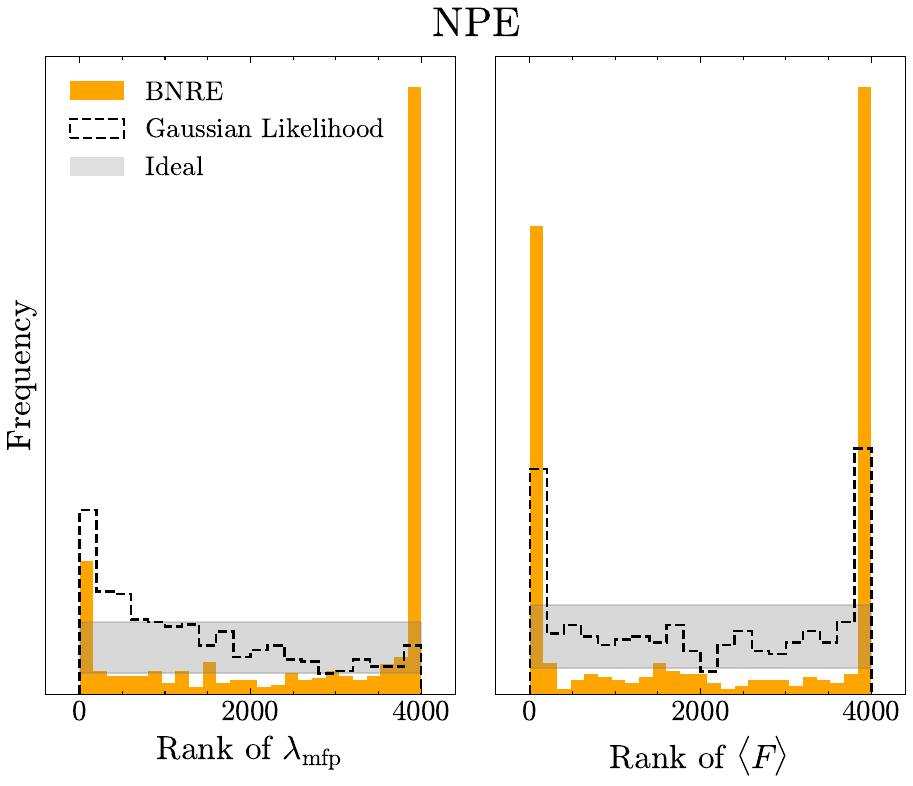}
    \hspace{0.07\textwidth} 
    \includegraphics[width=0.4\columnwidth]{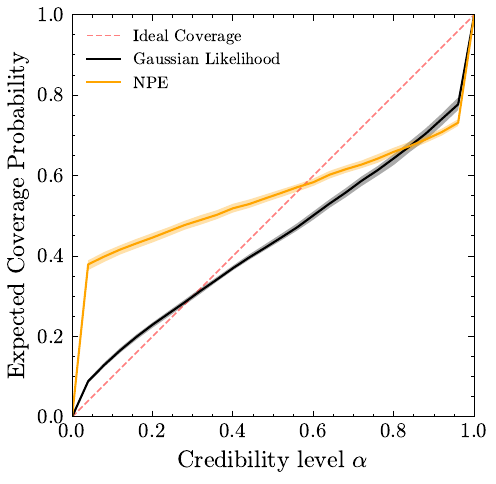}
    \caption{\textit{Left:} SBC rank histograms for $\lambda_{\text{mfp}}$ and $\langle F \rangle$ obtained NPE. Uniform posteriors indicate correct calibration, with the shaded region showing the expected range under sampling variability. \textit{Right:} Coverage probabilities obtained by using TARP on a set of posterior distributions obtained with both methods. The shaded regions represent the respective 16th--84th percentile ranges obtained via bootstrap sampling.}
    \label{fig:inference_tests_npe}
\end{figure*}

To evaluate its performance, we use the trained density estimator to infer posterior distributions for the same 500 mock observations as in Section~\ref{sec:results}, drawing 4000 samples per mock. Figure~\ref{fig:inference_tests_npe} shows the results of the TARP and SBC tests. 


The rank distributions for both parameters exhibit a strong U-shape, indicating that the trained NPE model is overconfident, consistent with the shape of the corresponding coverage curve. This overconfidence a well-known issue in some standard SBI algorithms \citep{Hermans2021}. However, we strongly emphasize that this result is not representative of NPE’s overall potential. We expect that with appropriate hyperparameter tuning or the use of a balanced objective \citep[as proposed by][]{Delaunoy2023}, NPE could likely achieve substantially better calibration.

\end{document}